\def\parder #1;#2;{{\partial #1\over \partial #2}}
\newcommand\ges{\gtrsim}
\newcommand\les{\lesssim}

\newcommand\etal{{\it et al.\ }}
\newcommand\refind{\hbox to 0.5 truein {\hrulefill .}}
\newcommand\beq{\begin{equation}}
\newcommand\eeq{\end{equation}}

\documentstyle[11pt,aaspp4]{article}

\begin{document}

\title{Interpretation of the   Spatial Power Spectra of Neutral Hydrogen
  in the Galaxy and in the Small Magellanic Cloud}
\righthead{The   Spatial Power Spectra of Neutral Hydrogen}
\author{Itzhak Goldman}
\affil{School of Physics and Astronomy, Sackler Faculty of Exact
  Sciences, Tel Aviv University, Tel Aviv 69978, Israel, email:
goldman@wise1.tau.ac.il}

\abstract

Recent  21 cm radio  observations of H$_I$ regions in the Small  Magellanic Cloud,
 have revealed   spatial power spectra of the intensity, which are quite
similar in shape 
to those previously deduced for
the Galaxy.
The similarity, in spite the differences in the physical parameters
between the Galaxy and the SMC,
suggests that 
the shape of the power spectra reflects some underlying mechanism
which is not too sensitive to the   environmental specifics.
In this paper
we present  an interpretation for the observational power spectra  in terms of
a large scale
 turbulence in the interstellar medium, in which the emitting H$_I$
 regions are embedded. 
The turbulence
gives rise to density fluctuations which lead to the observed  
intensity fluctuations, in the H$_I$
 regions.
The observational power spectra are  used to deduce  the turbulence
spectral function. 
 In the SMC, the turbulence largest eddies are comparable in scale to
the SMC itself. This implies that turbulent mixing should have
smoothed out 
any large scale abundance gradients. Indeed, this seems to be the case,
observationally. 

The turbulence is also expected to amplify and shape up the large scale magnetic field.
Indeed, the observational data indicate the existence of a large 
scale disordered field of the strength expected from energy
equilibrium with the turbulent  velocity field.
The large scale turbulence is most probably generated by instabilities 
in 
 the large scale  flows induced by the
tidal close encounter with the LMC $ \sim 2\times 10^8{\rm yr}$ ago.
The life-time of the largest eddies is
$\sim 4\times 10^8{\rm yr}$ so the turbulence had not yet enough time
to decay and persists even though the energy source is no longer there.


\section{Introduction}

Crovisier \& Dickey
(1983), and  Green (1993) have used 
21 cm radio observations to derive
the intensity  spatial power spectra for our Galaxy. Recently, Stanimirovi$\acute{{\rm c}}$
\etal (1999) obtained  H$_I$ spatial power spectra
 for the Small Magellanic Cloud (SMC).
Some  of the observations were  interferometric
and  yielded the power spectrum directly, and some  single dish
observations
from which the power spectrum  was calculated by Fourier transform.

The resulting  power spectra  exhibit  a power-law behavior,
$p(q)\propto q^{\gamma}$,
over a wide range of
spatial scales: $\sim 10$--$200{\rm pc}$ in the Galaxy, and $\sim 30{\rm pc}$--$4{\rm
  kpc}$ for the SMC. Here  $q$ is the absolute value of the
2-dimensional wavenumber in the plane of the sky, and 
the observational power-law indices are    $\gamma \sim  -(2.8\div  3) $ for the Galaxy
and   $\gamma \sim -3$ for the SMC.

Given the differences in the spatial scales for which the spectra were 
obtained,  and the differences in the physical parameters
between the Galaxy and the SMC,  the  similarity of the
indices is remarkable. It suggests the
existence of some underlying mechanism that is  not sensitive to the
 specific physical parameters of the environment. 
 Stanimirovi$\acute{{\rm c}}$
\etal  (1999) attributed the power-law shape to a fractal structure of
H$_I$ clouds hierarchy.

The observational power spectra reveal the existence of
intensity correlations over a wide range of spatial scales, and this  naturally
suggests  that turbulence in the interstellar medium (ISM) is the sought for
underlying mechanism.  Such an approach was suggested by Lazarian
(1995) and followed up by Lazarian \& Pogosyan (2000).
In this paper we present an interpretation of
the observational power spectra in terms of density fluctuations
 which are generated by a large scale 
turbulence in the ISM, in which the emitting neutral hydrogen regions are
embedded. 
The model determines the power spectrum for any given spectral function 
of the underlying   large scale turbulence.
Therefore, the  observational  power spectra can be used to study
 the above large scale turbulence. From  the observational
 power spectra, we deduce the  spectral function of the underlying
 turbulence:  $F(k)\propto k ^{-2}$, with $k$  denoting the absolute 
 value of the 3-dimensional wavenumber. As detailed in the Discussion 
 section, this spectral function is
quite common in various components of the ISM and characterizes  turbulence in a compressible medium.

\section{The Model Outline}

The observed lines of sight are generally optically thin
in the 21 cm line.  For the  SMC, 
 Stanimirovi$\acute{{\rm c}}$  \etal (1999) note that self absorption effects are
limited to narrow ranges in velocities and directions.  They find that  for
most of the  covered area the 
emission is from regions with column densities, integrated over the
entire velocity span, of
$N_{H_I}<2.5 \times 10^{21} cm^{-2}$. Each of the power spectra
corresponds to a velocity range which equals $1/6$ of the entire velocity
span. Thus, one can estimate that typically 
each power spectrum corresponds to a column 
density $N_{H_I}\sim 4 \times 10^{20} cm^{-2}$.

For the outer arms of the Galaxy, for which Green (1993) obtained  the
best fitting power spectra, we can estimate the column density using 
the relation (Spitzer 1978)

$$ N_{H_1}=1.82 \times 10^{18} cm^{-2} \int T_B dv \eqno(1)$$
where $T_B$ and $v$ are in $K$ and $km/s$, respectively. 
For the outer arms,  the average $T_B\sim 15 K$ over  velocity groups of 3
channels, each of width $ 2.64 km/s$ (Green 1993), yields  $N_{H_I}\sim 2\times 10^{20} cm^{-2}$.

To estimate the physical depth $D$ of the emitting neutral hydrogen
region, contributing to each power spectrum,
we take a
number density of neutral hydrogen of $n=10
{\rm cm}^{-3}$. This value is  intermediate
between $n \sim 5 {\rm cm}^{-3}$ corresponding to warm neutral hydrogen
with temperature $T\les
200{\rm K}$ , and $ n\ges 30{\rm cm}^{-3}$  of the cool neutral hydrogen with
$T\les 100 {\rm K}$ (Wolfire \etal 1995).
 This translates for the Galaxy outer spiral arms to a physical depth
 $D\sim 7 pc$ and for the SMC to $D\sim 15 pc$.

   The power spectra were grouped by the observers in  radial
velocity intervals. It is interesting that, both for the Galaxy and the
SMC, the 
power spectra for the different velocity
groups are similar.
In the Galaxy, adopting a rotation curve to deduce 
distances, each such velocity group represents a range of distances. 
Thus, in Green (1993)  this range for the
outermost spiral arms is $ \simeq 2.2{\rm kpc}$ while for the Perseus
data the range is $ \simeq 0.6 {\rm kpc}$.
Each such  distance range is {\it much larger} than  the
above estimate of $D \sim 7 pc$.   Therefore, it follows  that
the intensity is contributed by 
{\it disjoint subregions} of neutral hydrogen  with total depth  
equaling $D$, with each subregion
having a depth which is only a fraction of $D$, e.~g.  $\sim 1 pc$.

In the SMC, the  total velocity span probably represents large
scale flows, of few kpc, that are the 
result of tidal interactions with the LMC
and with the Galaxy. Indeed, N-body simulations  by Gardiner
\& Noguch (1996) indicate 
a very strong tidal impact imparted to the SMC by the LMC about $\sim 2\times
10^8 {\rm yr}$ ago, with a duration of $\sim 1\times 10^8$ yr. The
resulting  simulated velocity field 
exhibits velocities in the range of tens of km/s. This is 
consistent with the observational velocity span of $\sim 80 {\rm km/s}$
of  the neutral hydrogen emission as reported by Stanimirovi$\acute{{\rm c}}$
\etal (1999).
The  velocity range of each of the power spectra,
 corresponds to a distance interval $\sim 1$kpc which is {\it much larger} than
 the total depth of the emitting regions,  $D\sim 15 pc$ contributing
 to each power spectrum. Therefore in this case too, the emission
arises from a number of disjoint zones, each of depth  which is a fraction of
$D$, e. g. $\sim 2 pc$.

We note that in both cases, the depth of each of the disjoint emitting regions
 is also {\it much smaller}  than  the
 scales, in the plane of the sky,  probed in the power spectra: for the Galaxy  up to
$\sim 200$ pc and for the SMC
up to 4 kpc. Therefore, the  emitting H$_I$ regions occupy a small
fraction of  the ISM volume, in which they are embedded. A small
 space filling factor of neutral hydrogen is implied 
also in the fractal interpretation of  Stanimirovi$\acute{{\rm c}}$
\etal (1999).

 In the optically thin case, the intensity is
 proportional to the column density of neutral hydrogen in the line of  sight (Spitzer
 1978). Thus, the intensity fluctuations are due to density
  fluctuations in the disjoint neutral hydrogen regions. In the
  present model,
the density fluctuations are regarded to be a response to a large
 scale  3-dimensional turbulence in
the ISM, in which the neutral hydrogen regions  are embedded.

Since the observational optical depths are small, the following
picture  suggests itself : the $H_I$ subregions (clouds) are waved by a
large scale turbulence in the embedding medium. In this picture
the neutral hydrogen regions  are regarded as ``passive markers'' which follow the turbulence
field but do not feed back on it dynamically. The turbulence induces density fluctuations
in the neutral hydrogen subregions  that
manifest as intensity fluctuations.

To be specific,
 we consider a simple model  of
  density fluctuations in
thin   slabs or sheets parallel to the plane of the sky, which are
separated along the line of sight direction by distances 
{\it  much larger} than their depth. 
  This geometry is consistent with the above estimates of the depth of 
  the emitting subregions.
 H$_{\rm I}$ regions are generally assumed to have a filamentary
 structure and filaments are expected to 
have random orientations. This is not to say that all
 $H_I$ emitting regions are necessarily filamentary. However, in our case
 the contributions to the power spectrum  from a given angular 
 separation in the plane of the sky will
 predominantly be those from regions at roughly the {\it same}  
 distance along the line of sight. So,
 instead of summing over the actual filaments, contributing to each
 power spectrum,  we effectively sum over
 thin slabs parallel to the plane of the sky and  separated by
 distances large compared to their depth. One may regard the slab
 family as mathematically spanning the contributions of the actual
 physical filaments.

 For low galactic latitudes, where most of the neutral hydrogen is located,  there may
 indeed be a preference for   filaments alignment in the above thin slab geometry,
 as  a result of  the galactic differential rotation. In the
 SMC the rotation with respect to the Galaxy is expected to have a smaller effect.

In the next section, we show that the observational power spectra are
essentially the sum of the spectra contributed by each such
thin slab.  The  mutual contributions coming from 
{\it different} slabs
are much smaller.

\section{H$_{\rm I}$ Spatial  Power Spectra and ISM Turbulence}

We wish to relate the observational two-dimensional power spectrum with
the underlying three-dimensional turbulence. 
 This will provide  an
interpretation of the observational power spectrum and at the same time
could hopefully yield clues regarding the  
characteristics  of the underlying 3-dimensional turbulence.

The observational
2-dimensional power spectrum is  defined as the Fourier transform, over
the plane of the sky, of the autocorrelation of the fluctuating
 brightness
temperature $\theta_B=T_B-<T_B>$.
Here $T_B$ is the brightness temperature and $<T_B>$  its 
 average  over the plane of the sky. Thus, the power spectrum $p(\vec{q})$
as function of the 2-dimensional wavenumber in the plane of
the sky, $\vec{q}$,  is

$$p(\vec{q})= \frac{1}{(2\pi)^2}\int  <\theta_B(\vec{R}) \theta_B(\vec{R}+\vec{r})> 
 e^{-i \vec{q}\cdot \vec{r}} d^2 r \  .\eqno(2)$$
Here  $\vec{R}$,  and $\vec{r}$ are in the plane of the sky,
and
$<\theta_B(\vec{R}) \theta_B(\vec{R}+\vec{r})> $ is the
autocorrelation  of $\theta_B$ obtained by
 averaging over all positions $\vec{R}$ in the plane of the
sky.  Assuming homogeneity and isotropy in this plane,  the power
spectrum depends only on  the absolute value of the wavenumber, so
that $p(\vec{q})=p(q)$ with $q=|\vec{q}|$.
 In the optically thin case, the brightness temperature is
 proportional to the 
column density of neutral hydrogen in the line of
 sight (Spitzer 1978), thus in the present case

$$T_B(\vec{R})=B N_{\rm H _I}(\vec{R})= B\sum_i \int_{d_i-L_i/2}^{d_i+L_i/2}
n(\vec{R},z) dz \eqno(3)$$
with $n$ the number density and $B$ a constant.
The summation is over the disjoint thin slabs
with depth $L_i $,  located at   distances $d_i$. As mentioned before, 
each depth $L_i$
 is {\it much smaller} than the slab extent on
the plan of the sky 
 and also {\it much smaller} than the distances between slabs.
The fluctuating part of the brightness temperature is thus expressed as

$$\theta_B(\vec{R})= B\sum_i \int_{d_i-L_i/2}^{d_i+L_i/2}\tilde n(\vec{R},z) dz \eqno(4)$$
where
$\tilde n=n-<n>$
is the fluctuating number density,  and $ <n>$ is the average number
density in the slab.
The autocorrelation of $\theta_B$ is thus

$$<\theta_B(\vec{R}) \theta_B(\vec{R}+\vec{r})>=B^2 \sum_{i,
  j}\int_{d_i-L_i/2}^{d_i+L_i/2}\int_{d_j-L_j/2}^{d_j+L_j/2} <\tilde n(\vec{R},z')
\tilde n(\vec{R}+\vec{r}, z)> dz dz'\eqno(5)$$
where, again,  the angular brackets represent averaging over all positions 
$\vec{R}$ in the plane of 
the sky.

From Crovisier \& Dickey
(1983), Green (1993), and  Stanimirovi$\acute{{\rm c}}$
\etal (1999)
it can be estimated that, typically, the root mean square of
$\theta_B$, $\theta_{B rms}\simeq  (0.1 \div 0.2)
<T_B>$ implying that the root mean square of $\tilde n$,   
$\tilde n_{rms}\simeq (0.1\div 0.2) <n>$. Thus, although
the medium is compressible,  the density fluctuations giving rise
to the brightness temperature fluctuations are not large.
This renders 
plausible the assumption that the density 
fluctuations are  driven by a velocity turbulence and are actually
a ``passive scalar'', i. e. trace the velocity fluctuations but {\it do not
feed back} upon them.  
In this case, the 2 -point correlation
of a density fluctuations field 
  driven by a turbulent velocity field,  is proportional to the 2-point 
  correlation of the latter (see e.g. Lesieur 1997)

$$ <\tilde n(\vec{R},z') \tilde n(\vec{R}+\vec{r},z)>=A^2
<\vec{v}(\vec{R},z')\cdot \vec{v}(\vec{R}+\vec{r}, z)>\eqno(6)$$
with  $A$ a constant.
For simplicity we further assume that the 
underlying 3-dimensional turbulence is  homogeneous and isotropic
so 
that the plane average  equals the ensemble average 

$$<\vec{v}(\vec{R},z')\cdot \vec{v}(\vec{R}+\vec{r}, z)>=C_3(\vec{r}, z-z')\eqno(7)$$
where $C_3$ is the  2-point
correlation of  the underlying 3-dimensional
 velocity turbulence,  defined in terms of {\it ensemble} average

$$C_3(\vec{r})=<\vec{v}(\vec{r'}) \cdot \vec{v}(\vec{r}+\vec{r'})>=\int
\Phi(\vec{k})e^{i \vec{k}\cdot \vec{r}}  d^3k\eqno(8)$$
In equation (8), unlike in the previous equations, the position vectors
are 3-dimensional  and 
the brackets denote
  {\it ensemble} average. $C_3$ is expressed in
terms of the 3-dimensional spectrum $\Phi(\vec{k})$ which in the
homogeneous and isotropic case depends on   $k$, the absolute value of
the 3-dimensional wavenumber $\vec{k}$,

$$\Phi(\vec{k})=\Phi(k)=\frac{ F(k)}{4\pi k^2}\ . \eqno(9)$$
Here, $F(k)$ is 
the
 spectral function of the 3-dimensional velocity turbulence. It equals
 twice the kinetic energy per unit mass and unit wavenumber in the turbulence. 
 Accordingly, the mean squared  value of the turbulent velocity  is 
$$v_t^2=\int _{k_0}^{\infty }F(k)dk\eqno(10)$$
with $\pi k_0^{-1}$ being the largest scale in the turbulence.

 We can now go back to equation (2) and use equations (5), 
(6), (8) and (9)
 to
 express $p(q)$ in terms of
$F(k)$

$$p(\vec{q})= \frac{A^2B^2}{(2\pi)^2}\sum_{i,
  j}\int_{d_i-L_i/2}^{d_i + L_i/2}\int_{d_j -L_j/2}^{d_j +L_j/2}\int
\Phi(\vec{q'}, k_z)e^{i \vec{q'}\cdot \vec{r} -i \vec{q}\cdot
  \vec{r}+ik_z (z-z')} dk_z\  d^2q'\   d^2 r\ 
dz\  dz'\ ,\eqno(11)$$
leading to

$$p(\vec{q})=A^2B^2 \int
\Phi(\vec{q}, k_z)\left( \sum_{i,
  j}\int_{d_i -L_i/2}^{d_i +L_i/2}\int_{d_j -L_j/2}^{d_j +L_j/2}e^{ik_z (z-z')} 
dz dz'\right) dk_z=\eqno(12)$$

$$A^2B^2\int \Phi(\vec{q}, k_z)\left( \sum_i L_i^2 \frac{{\rm
      sin}^2 (k_zL_i/2)}{(k_zL_i/2)^2}+\sum_{i\neq j} L_iL_j e^{ik_z(d_i-d_j)}
\frac{{\rm  sin} (k_zL_i/2)}{k_zL_i/2} \frac{{\rm sin}
  (k_zL_j/2)}{k_zL_j/2}\right) dk_z\ .$$
 Since $|d_i-d_j | \gg L_i, L_j $,  the contribution of the second term which
results from inter-slab correlations is {\it much smaller} than that of the
first term that is due to correlations {\it within} each slab. Keeping only
the first term gives

$$p(\vec{q})=A^2B^2\sum_i L_i^2  \int_0 ^\infty\Phi(\vec{q}, k_z)\frac{{\rm
      sin}^2 (k_zL_i/2)}{(k_zL_i/2)^2} dk_z\ .\eqno(13)$$
The assumed homogeneity and isotropy in the plane implies that
$p(\vec{q})=p(q)$, and use of equation (9) leads to

$$p(q) =\frac{A^2 B^2}{4\pi}\sum_i L_i^2  \int_0^\infty
\frac{F\left(\sqrt{q^2 +k_z^2}\ \right)}{(q^2 +k_z^2)}\frac{{\rm  sin}^2
  (k_zL_i/2)}{(k_zL_i/2)^2} dk_z\eqno(14)$$
with  $q$ being  the absolute value of the 2-dimensional in-plane
wavenumber, $\vec{q}$. Here $k_z$ denotes the wavenumber in the $z$ direction,
perpendicular to the plane of the sky, so that $k^2=q^2+k_z^2$.
The  power spectrum is thus the sum of the power spectra due to each of
the thin slabs. Equation (14) determines the power spectrum $p(q)$
for any given spectral function $F(k)$ of  the
underlying 3-dimensional turbulence .

\section{Application to the Observational Power Spectra}

It is of interest to use equation (14) in order to deduce information
about $F(k)$ from the observational  power spectrum $p(q)$. In the general
case this cannot be done uniquely. Nevertheless, for power-law
functions it is possible.

Since 
 the depth of each slab is much smaller than the observed
scales in the plane of the sky  for
which the power spectra were  obtained,   the wavenumbers  satisfy
$q<<\pi/L_i$. From equation (14) follows that  in this case, for
an  underlying 3-dimensional turbulence  spectrum which is a 
power- law,  $F(k)\propto k^{-m}$, 
the  2-dimensional power spectrum is also a
power-law, $p(q)\propto q^{-m-1}$.
To see this,  note that each of the integrals in equation (14) is now  of
the form
$$ \int_0^{\infty}\left (q^2+k_z^2\right)^{-m/2-1}\frac{{\rm  sin}^2
  (k_zL_i/2)}{(k_zL_i/2)^2} dk_z= I(q) q^{-m-1}\eqno(15)$$
where

$$I(q)=\int_0^{\infty} \left (1+y^2\right)^{-m/2-1} \frac{{\rm  sin}^2
  ( y qL_i/2)}{(y q L_i/2)^2}dy\eqno(16)$$
and  $y= k_z/q$.

 A
plot  of  $I(q)$ is presented in figure 1
  for  three representative values of $m=1, 2, 3$. The wavenumber $q$ is in units of
$q_0$, 
corresponding to the largest scale in the plane of the sky. A value
 $q_0 L_i/\pi=0.005$  typical
to the Galaxy was adopted; for the SMC a typical value
is $0.002$.  
For each $m$,
$I(q)$ is essentially constant down to
scales in 
the plane of the sky that 
are $\sim 100$ times smaller than the largest scale. Only for very small scales, which turn to be 
smaller than the observed ones,   $I(q)\propto q^{-1}$. For the SMC value
of  $q_0 L_i/\pi$,
 the range of scales for which  $I(q)$  is constant, would be  wider.
  Thus, each of the
terms in equation (14) is a power-law with the {\it same}
index $-(m+1)$.

The above results can be understood qualitatively quite easily.
The  term ${\rm  sin}^2
  (k_zL_i/2)/(k_zL_i/2)^2$ effectively limits the integration
 over $k_z$ to the range $\sim 0\div \pi/L_i$. Approximating it
by its average value, yields
 
$$ I(q)\propto \int_0^{\pi/(L_iq)} \left (1+y^2\right)^{-m/2-1} dy\eqno(17)$$
Indeed, in the limit of $q<<\pi/L_i$ which applies here, $I(q)$ is constant.

The power spectrum for each of the velocity groups is a sum of  the spectra
 of the individual slabs. Since
each of these  power spectra is a power-law  with the {\it same} index, the
combined power spectrum will also be a power-law with this index.
We conclude that a power spectrum $p(q) \propto q^{-m-1}$ is the
result of an underlying 3-dimensional turbulence characterized by a spectral
function  $F(k)\propto k^{-m}$.
The observational values of the index of the power spectrum, 
imply that 
  $ m\sim 1.8\div 2$ for the Galaxy
and $m\sim 2$ for the SMC.

\begin{figure}[h]
  \begin{center}
    \leavevmode
    \plotone{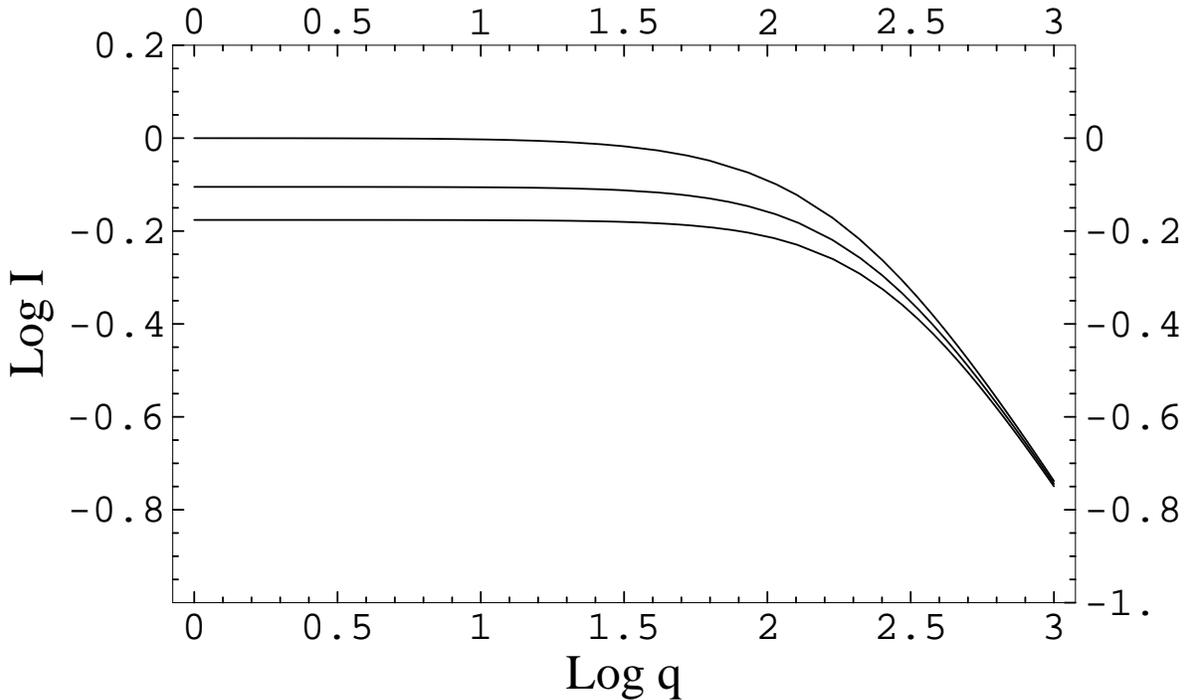}
\figcaption
{Log $I(q)$ as function of Log $q$, (in units of $q_0$), for $q_0
  L_i/\pi=0.005$. The curves correspond to $m=1, 2, 3$ in this order,
  with $m=1$ being the uppermost one.}

\end{center}
\end{figure}

\section{Discussion}

To interpret the observational  spatial power spectra
 we have assumed 
   that the
 density fluctuations that produce them 
  are a response to a large
 scale  3-dimensional turbulence  in
the ISM, in which the emitting neutral hydrogen  regions are embedded.
We derived an expression for the intensity power spectrum in terms of
the underlying 3-dimensional large scale turbulence. 

The power spectra 
were seen to be  the sum of contributions from thin slabs of neutral
hydrogen, parallel to the plane of the sky,
with depth {\it  much smaller} than their extent in the plane of the sky. The 
depth is also {\it much smaller} than the distances, along the line of sight,
between slabs. In this case, we found that a power-law 3-dimensional turbulence
spectrum, $F(k)\propto k^{-m}$, leads to a power-law 2-dimensional
intensity  power spectrum,
$p(q)\propto q^{-m-1}$. From the observational values of the power spectrum 
indices, it follows that $m\sim 2$ for the SMC and $m=1.8 -2$ for the Galaxy.

\subsection{Turbulence spectrum}

This 3-dimensional turbulence spectrum resembles  the Kolmogorov spectrum 
characterized by  $m=5/3$ but differs in the value of the index which
is closer to $m=2$. For the  Kolmogorov spectrum the turbulent
velocity on scale $l$ satisfies $v_t(l)\propto l^{1/3}$, while in the present
case $v_t(l)\propto l^{1/2}$.
A velocity-size relation  with index $\sim 0.5$,
has been deduced for  molecular
clouds of  a wide range of sizes 
(Larson 1981; Leung,
Kutner \& Mead 1982; Myers 1983; Dame \etal 1986; Myers \& Goodman 1988;
Falgarone, Puget \& Perault 1992; Miesch \& Bally
1994). A turbulence spectrum characterized by $m=2$ was also 
found  in a H$_{II}$ region by Roy \& Joncas (1985).
 It is interesting that this 
 spectrum seems to prevail in different components of the ISM. 
The common factor in these cases is that the medium is compressible.
Indeed,   numerical simulations, with and without a magnetic field
 (Passot, Pouquet,
\& Woodward 1988, V$\acute{{\rm a}}$zquez-Samedani, Ballesteros-Paredes, \&
  Rodrigu$\acute{{\rm e}}$z 1997), indicate that 
such a
 spectrum is characteristic to turbulence in a compressible medium
 (whether or not magnetic field fluctuations are important).

 The Kolmogorov spectrum strictly applies 
to the inertial wavenumber range for homogeneous and isotropic incompressible
turbulence. In this range there is no energy input from the source
generating the turbulence (input is at larger scales)  nor energy
losses (dissipation is at smaller scales). Thus, turbulent  kinetic  energy is cascaded 
from larger to smaller scales due to the nonlinear eddy interactions, 
at a {\it constant rate} that is {\it independent of
wavenumber}.

The present steeper spectrum, with $m\sim 2$, indicates that at each
wavenumber a
part  of the cascaded turbulent kinetic energy is transformed to other energy
forms. As a result, the turbulent kinetic energy rate is smaller the larger the  wavenumber.
In a  compressible medium, part of the kinetic turbulent energy is indeed
transformed to energy associated with  density fluctuations.

Finally, we note that the emitting neutral regions themselves, modeled 
here as thin slabs, are
probably transient features created and disrupted by the large scale 
turbulence (Ballesteros-Paredes, V$\acute{{\rm a}}$zquez-Samedani, \&
Scalo 1999). Their life-times are of the order of the
timescales of the largest eddies. The spatial scales for which
the  power
spectra were obtained are smaller than the above, and therefore the corresponding  turbulence timescales are
shorter than those of the very largest scales. With respect to the
observed scales, the slab can be considered as a
quasi-stationary
structure in which the observed smaller scale eddies generate the density
fluctuations that give rise to the intensity fluctuations.

\subsection{Turbulence Scales and Energy Sources}

The scale of the large-scale turbulence in the Galaxy should  
exceed 200 pc -- the largest scale probed by the power spectra. The
power-law shape of the spectra suggests that energy input to the
turbulence is on yet
larger scales.
A probable  energy source on these larger scales is the
galactic  differential rotation, as also suggested by numerical simulations of Wada 
\& Norman
(1999).
In this case, a natural  scale will be a fraction of the galactocentric
distance. However, for  a 3- dimensional turbulence, the
largest  scale is likely to be comparable to the width of
the galactic  H$_{II}$ layer 
 of  $\sim$~2kpc (Reynolds 1989; Reynolds 
1991). 
 
 In the SMC, the turbulence can be 
 generated by a host of instabilities related to the bulk flows that
 result from the tidal interaction with the Galaxy and with the LMC,
 as well as to the expanding shells and super shells. The existence of
 large
 scale correlated density fluctuations in the
SMC indicates the existence of turbulence on a scale $\ges$~4 kpc which is
comparable to the size of the SMC . This 
observational evidence by
itself is  a  very interesting result as the turbulence scale is an order of magnitude
larger than  that of the turbulence in the LMC,  studied by Spicker \&
Feitzinger (1988) on the basis of radial velocities of 21-cm emission.

The most plausible source feeding energy to such an extremely large scale turbulence 
are bulk flows induced by the tidal interactions with the Galaxy and the
LMC, notably the very close encounter with the LMC that peaked about $2\times
10^8{\rm yr}$ ago and lasted for about $2\times
10^8{\rm yr}$  (Gardiner \& Noguchi 1996). Such large scale bulk
flows 
 generate turbulence due to shear instabilities e. g. the
Kelvin-Helmholtz instability (KH) or  shock induced Richtmier-
Meshkov instability(RM). 
These two mechanisms were shown to induce large scale turbulence in
the intracluster gas  of merging galaxy clusters by Roettiger \etal (1998) and by Goldman
(1997), respectively.

The  simulations of  Gardiner
\& Noguchi (1996)  exhibit large scale (few
kpc) velocity gradients with shear values  $S\simeq  5-10{\rm
  km/s/kpc}$,
consistent with the (long period, population I) Cepheid radial velocity data of Mathewson, Ford,\& Visvanathan (1986, 1988). One  expects
that the large scale shear will produce turbulence with the largest
eddies size  $l_t$ comparable to these spatial scales, and root mean
squared  turbulent
velocities $v_t\les S l_t=10\div 15{\rm km/s}$. This value is 
comparable to the sound speed of
the embedding ionized medium, and supersonic with respect to the H$_I$ 
regions, thus enabling the density fluctuations in the latter.
 On  smaller scales of $\les$~1~kpc, 
also the giant super-shells observed in the
SMC, contribute, to the turbulence. The giant supersehlls themselves
are probably the result of a star formation burst that followed the
interaction with the LMC (Zaritsky \etal 2000).

The time scale for decay of the  turbulence $t_d\sim l_t/v_t$ with a  rms
turbulent velocity  of
$v_t\approx 10 {\rm km/s}$ and scale $l_t=1 {\rm kpc}$ is $\sim 4 \times
10^8{\rm yr}$. Thus, we are witnessing a turbulence that didn't have enough
time to decay, since the close encounter with the LMC of $\sim 2\times 10^8{\rm yr}$ ago.
Moreover, the simulations of Roettiger \etal (1998) and simulations of
RM instability (Rotman, 1991) indicate that continued excitation (even at a lower
level than the original one) can keep such a turbulence going for times considerably
longer than the above estimate. Such a lower level excitation can be
provided by ongoing tidal interactions of smaller' strength.

Another question of interest is the near constancy of the slope of the 
spectra. We know that small scale (from 1kpc and down) ) energy
sources exist (supergiant shells, smaller scale supernovae
remnants, smaller scale hydrodynamic instabilities) as well as
smaller scale energy sinks (e.g. magnetic excitations). Consider
  a source /sink a  at wavenumber $k$ in an
interval $\Delta k$ such that $k\ges \Delta k>> k_0$ with $k_0$
corresponding to the largest scale in the turbulence. Denote 
by  $\epsilon$  and $\Delta \epsilon$   the energy rate due to the 
large scale turbulence and the
change due to the  small scale source, respectively.
Since  $F(k)\propto \epsilon(k)^{2/3}$ the
jump in the turbulence spectral function 

$$\frac{\Delta F(k)}{F(k)}\sim \frac{2}{3}\frac{\Delta
  \epsilon}{\epsilon}\ . \eqno(18)$$
Even for a relative change of $\sim 20\%$, the jump will be $\les
10\%$,  consistent with the scatter  of the
observational spectra.

\subsection{Implications of the Turbulence on the SMC Magnetic Field
  and Abundances Gradients}

The consequences of this turbulence may be quite important regarding
magnification of large scale magnetic fields, and enhancing diffusion
coefficients. 
In the SMC, the turbulence largest eddies are comparable in scale to
the size of the SMC. This implies that turbulent mixing should smooth
any large scale abundance gradients. This seems indeed to
be the case observationally (Dufour 1975; Kobulnicky 1998) .
The turbulence is also expected to amplify and shape up the large scale magnetic field.
Indeed, the SMC observational data indicate the existence of a large 
scale disordered field  (Haynes  \etal 1991; Ye \& Turtle 1991). The
observed 
magnetic field is $\sim 6.5\mu{\rm G}$. a value close to that expected
from a 
dynamo amplification to an  equilibrium
value  $B=5\mu{\rm G}
(v_t/10{\rm km s}^{-1})(n/1 {\rm cm}^{-3})^{1/2}$ with $v_t$ the turbulent
root mean squared velocity and $n$  the 
number 
density in the ionized embedding region.

\subsection{Concluding Remarks}

The interpretation of the $H_I$ power spectrum of the SMC in terms of an
underlying large scale turbulence seems natural and as noted above is
consistent with other observational data.
The existence of a large scale turbulence in the SMC gains also
support from the recent finding of Stanimirovi$\acute{{\rm c}}$ \etal
(2000), that the autocorrelation of the dust column
density is also a power law with an index $\sim -3$. In our
interpretation, the dust will be just another passive marker waved by
the turbulence and thus should have the same power-law spectral index 
as the $H_I$ intensity.

Moreover,  observations of stars of various ages were used  by Zaritsky \etal (2000) and by
Kunkel, Demers, \& Irwinl (2000), to conclude
that hydrodynamic interactions, following the recent close
passage of the LMC, rather than just gravity are responsible for the
morphology and dynamics of the ISM of the SMC.

\vskip 0.5 truecm

Comments by the referee and by Dr. Steven Shore have contributed to the
improvement of the presentation. 
This work has been supported by the  US-Israel BSF grant 94-314 and 
Israel Science Foundation grant 561-9911.2.


\begin{thebibliography}{}
\bibitem[]{} Ballesteros-Paredes, J., V$\acute{{\rm a}}$zquez-Samedani, E., \& Scalo,
  J. 1999. \apj, 515, 286
\bibitem[]{} Crovisier, J., \& Dickey, J. M. 1983, \aap, 122, 282
\bibitem[]{} Dame, T. M., Elemgreen, B. G., Cohen, R. S., Thaddeus,
P. 1986, \apj, 305, 892
\bibitem[]{} Dufour, R. J. 1975, \apj, 195, 315
\bibitem[]{} Falgarone, E., Puget, J.-L., \& Perault, M. 1992, \aap, 257, 715
\bibitem[]{} Gardiner, L. T., \& Noguchi, M. 1996, \mnras, 278,191
\bibitem[]{}Goldman, I. 1998, In Proceedings of Iau symposium 188 "The Hot Universe",
Kyoto, Japan The hot universe 
edited by Katsuji Koyama; Shunji Kitamoto; Masayuki Itoh. Dordrecht : Kluwer
                Sym. No.188., p.297
\bibitem[]{} Green, D.  A. 1993, \mnras, 262, 327
\bibitem[]{}  Haynes, R. F., Klein, U., Wayte, S.R., Wielebinski, R,
  Murray, J. D., Bajaaaaaja, E., Meinert,
  D., Buczilowski, Harnett, J.I., Hunt, A.J., Wark, R., \& Sciacca,
  L. 1991, \aap, 252, 475
\bibitem[]{} Kobulnicky, C. 1998, 
PASP Conference Series, 147,108
\bibitem[]{} Kunkel, W. E., Demers, S., \& Irwin, M. J. 2000, astro-ph/0003395
\bibitem[]{} Lazarian,  A. 1995, \aap, 293, 507 
\bibitem[]{} Lazarian,  A. \& Pogosayn, D. 2000, \apj, accepted
\bibitem[]{} Larson, R. B. 1981, \mnras,194, 809
\bibitem[]{} Lesiurs, M. 1997, Turbulence in Fluids, \S 6.10  (Dordrecht: Kluwer)
\bibitem[]{} Leung, C. M., Kutner, M. L., \& Mead, K. N. 1982, \apj,
  262, 583
\bibitem[]{} Mathewson, D. S., Ford, V. L., \& Visvanathan, N. 1986
  \apj, 301, 664
\bibitem[]{} Mathewson, D. S., Ford, V. L., \& Visvanathan, N. 1988
  \apj, 333, 617
\bibitem[]{} Miesch, M. S., \& Bally, J. 1994, \apj, 429, 645

\bibitem[]{} Myers, P. C. 1983, \apj, 270, 105
\bibitem[]{} Myers, P. C., \& Goodman, A. A. 1988, \apj, 329, 392
\bibitem[]{} Passot, T., Pouquet, A., \& Woodward, P. 1988, \aap, 197, 
  228
 
\bibitem[]{} Reynolds, R. J. 1989, \apj, 339, L29
\bibitem[]{} Reynolds, R. J. 1991, in The Interstellar Disk-Halo
  Connection in Galaxies, IAU Symp. No.144, ed. H. Bloemen (Dordrecht: 
  Kluwer), p. 67

\bibitem[]{} Roettiger, K., Stone, J. M., \& Mushotzky, R. F. 1998, \apj, 493, 62
\bibitem[]{} Rotman, D. 1991, Phys. Fl. A3, 1792;
\bibitem[]{} Roy, J. R., \& Joncas, G. 1985, \apj, 288, 142
\bibitem[]{} Spicker, J., \& Feitzinger, J. V. 1988, \aap, 191, 10
\bibitem[]{} Spitzer, L. Jr. 1978, Physical Processes in the
  Interstellar Medium (New York:Wiley)
\bibitem[]{} Stanimirovi$\acute{{\rm c}}$, S., Stavely-Smith, L.,
  Dickey, J. M., Sault, R. J., \& Snowden, S. L. 1999, \mnras, 302,
  417
\bibitem[]{} Stanimirovi$\acute{{\rm c}}$, S., Stavely-Smith, L.,
van der Hulst, J. M., Bontekoe, Tj. R., Kester, D. J. M., \& Jones, 
P. A. 2000, \mnras
 
\bibitem[]{} Ye, T., \& Turtle, A.J., 1991, \mnras,  249, 693
\bibitem[]{} V$\acute{{\rm a}}$zquez-Samedani, E., Ballesteros-Paredes, J., \&
  Rodrigu$\acute{{\rm e}}$z, L. F. 1997, \apj, 474, 292
\bibitem[]{}Wada, K., \& Norman, C. A. 1999, \apj, 516, L13

\bibitem[]{}Wolfire, M. G., Hollenbach, D., McKee, C. F., Tielens,
  A. G. G. M.,  \& Bakes, E. L. O. 1995, \apj, 443, 152
\bibitem[]{} Zaritsky, D., Harris, J., Grebel, E. K., \& Thompson,
  I. B. 2000, astro-ph/0003155

\end{thebibliography}
\end{document}